# Chemifriction and Superlubricity: Friends or Foes?

Penghua Ying[1], Xiang Gao[1], Amir Natan,[2] Michael Urbakh[1], Oded Hod[1]

[1] *Department of Physical Chemistry, School of Chemistry, The Raymond and Beverly Sackler Faculty of Exact Sciences and The Sackler Center for Computational Molecular and Materials Science, Tel Aviv University, Tel Aviv 6997801, Israel*
[2] *Department of Physical Electronics, Tel-Aviv University, Tel-Aviv, 6997801, Israel*

**Abstract:**

The mechanisms underlying chemifriction, i.e. the contribution of interfacial bonding to friction in defected twisted graphene interfaces are revealed using fully atomistic machine-learning molecular dynamics simulations. This involves stochastic events of consecutive bond formation and rupture, that are spatially separated but not necessarily independent. A unique shear-induced interlayer atomic transfer healing mechanism is discovered that can be harnessed to design a run-in procedure to restore superlubric sliding. This mechanism should be manifested as negative differential friction coefficients that are expected to emerge under moderate normal loads. A physically motivated phenomenological model is developed to predict the effects of chemifriction in experimentally relevant sliding velocity regimes. This allows us to identify a distinct transition between logarithmic increase and logarithmic decrease of frictional stress with increasing sliding velocity. While demonstrated for homogeneous graphene interfaces, a similar mechanism is expected to occur in other homogeneous or heterogeneous defected two-dimensional material interfaces.

**Introduction**

Sliding at the interface of two rigid, flat, and weakly interacting crystalline surfaces that are stacked in an incommensurate configuration, leads to ultra-low friction (1, 2). This is due to the effective cancellation of lateral forces opposing translational motion — a hallmark of a phenomenon referred to as structural superlubricity (3-8). Van der Waals (vdW) interfaces of layered contacts are ideal platforms to observe superlubricity, where interfacial incommensurability is facilitated by, e.g., the misfit angle in twisted graphitic interfaces (9, 10), curvature effects in multiwalled carbon nanotubes (11), intrinsic lattice mismatch in graphene/hexagonal boron nitride heterostructure (7, 12), and strain engineering in bilayer graphene (13, 14). The scaling up of superlubricity from the nanoscale to macroscopic contacts inevitably involves structural defects (15), such as atomic vacancies (16, 17), steps (18, 19), edges (20, 21), and grain boundaries (22, 23). The impact of such defects on the friction of superlubric sliding interfaces involves two distinct contributions: (i) physical contributions from defect-induced out-of-plane corrugation; and (ii) chemical contributions associated with the formation and rupture of interlayer covalent bonds. While the former have been extensively explored over the past decade (24-33), the atomistic insights into the latter remain much less understood. We recently demonstrated that this challenge can be addressed by machine-learning graph neural network potentials. Specifically, for bilayer defected graphene, we developed such a potential that achieves *ab initio* level accuracy in simultaneously describing sliding energy corrugation and interlayer bonding dynamics in superlubric layered contacts (34).

In the present letter, we use this computational framework to elucidate the phenomenon of kinetic *chemifriction* — the friction arising due to chemical bonding dynamics. We perform non-equilibrium molecular dynamics (MD) simulations to study the stochastic nature of interlayer bonding dynamics in defected twisted bilayer graphene, and the corresponding impact on structural superlubricity of this incommensurate sliding interface. Based on the numerical results we then construct a theoretical approach that allows us to bridge between atomistic simulations and experimentally relevant time- and length-scales.

**Methods and models**

Our model system consists of a 9.43° twisted defected graphene bilayer with a single vacancy in each layer (marked as $V_1V_1$), encompassing overall 590 atoms (see Fig. 1a). In accordance with experimental observations,(35-38) the single atomic vacancies are taken to be unpassivated in our model system. The two vacancies were positioned to ensure that upon relative sliding of 2.59 nm (half of the box length) between the top and bottom layers along the zigzag ($x$-axis) direction they align in an eclipsed configuration. Overall, the laterally periodic $V_1V_1$ model has a vacancy fraction of approximately 0.3%, comparable to experimental observations (17).

The reactive sliding dynamics simulations were performed using the recently developed machine learning potential (34), as implemented in the LAMMPS package (39). In these simulations, the top layer is driven laterally along the zigzag direction by a rigid slider (replicating the initial top layer) via a harmonic spring, whereas the bottom layer atoms are anchored to their initial positions with springs of the same stiffness (see Supporting Information (SI) Fig. S1). Due to the stochastic nature of the interlayer binding process (see, e.g., Ref. (40)), a minimum of 50 independent trajectories (each with a different random number seed for the initial atomic velocity distribution) was used to evaluate the bond formation probability and calculate the average kinetic friction. Unless noted otherwise, each trajectory involved a total sliding displacement of 5.18 nm, corresponding to the dimension of the supercell along the zigzag direction. To characterize interlayer bond formation and rupture, we used an interatomic distance criterion of 1.8 Å for atoms residing in different layers. More details regarding the reactive sliding dynamics simulation setup are provided in SI Section S1.

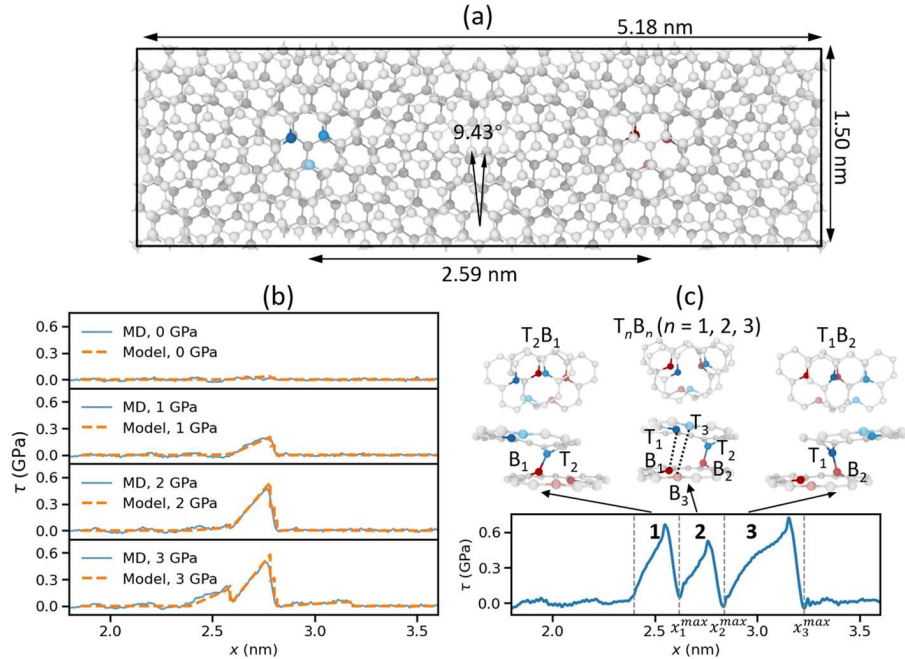

Fig. 1. $V_1V_1$ interface sliding dynamics simulations at a temperature of 300 K and a relative sliding velocity of 10 m/s. (a) Top view of the 9.43° twisted $V_1V_1$ bilayer graphene considered in this work. The unsaturated dangling atoms around the vacancies in the top and bottom layer are highlighted in blue and red, respectively. (b) Shear stress traces of the twisted $V_1V_1$ bilayer under external normal loads of 0, 1, 2, and 3 GPa obtained from the atomistic MD simulations (solid blue lines) and the developed phenomenological model (Eq. 8, dashed orange lines). (c) A representative trajectory demonstrating consecutive bond formation and rupture events in three regions (marked by 1-3 and enclosed by vertical dashed lines) occurring under a normal load of 3 GPa. The friction trace (bottom), and corresponding snapshots, showing binding pathways in top (upper row) and front (middle row) views, are presented. For clarity, only atoms in the vicinity of the defects are presented, with the six dangling atoms highlighted in blue (top layer) or red (bottom layer). Atomistic snapshots were visualized using the OVITO package (41).

**Results**

We first examined the impact of external normal load on the bonding dynamics and sliding friction. Figure 1b displays four representative friction traces obtained at a temperature of $T = 300$ K and a sliding velocity of $v = 10$ m/s, under external normal loads of $\sigma = 0, 1, 2,$ and $3$ GPa. Similar to previous results (34), we find that under zero normal load, no apparent covalent bonding between the two layers occurs even when the defects are eclipsed, leading to a relatively smooth shear trace and low friction. Increasing the load to 1 or 2 GPa results in the formation of covalent bonds, as manifested by the pronounced shear-stress peak appearing at a lateral displacement of $x \approx 2.75$ nm, slightly beyond the eclipsed defects configuration. At a higher normal load to 3 GPa, two additional shear-stress features appear (at $x \approx 2.55$ and $x \approx 3.15$ nm), corresponding to consecutive bond formation and rupture events (see SI Movies S1-S3 for single and multiple binding events). These results suggest that, beyond the increase of friction with normal load due to enhanced Pauli repulsions at the pristine interface regions, covalent bonding in defected regions introduces a significant positive contribution to the kinetic friction coefficient.

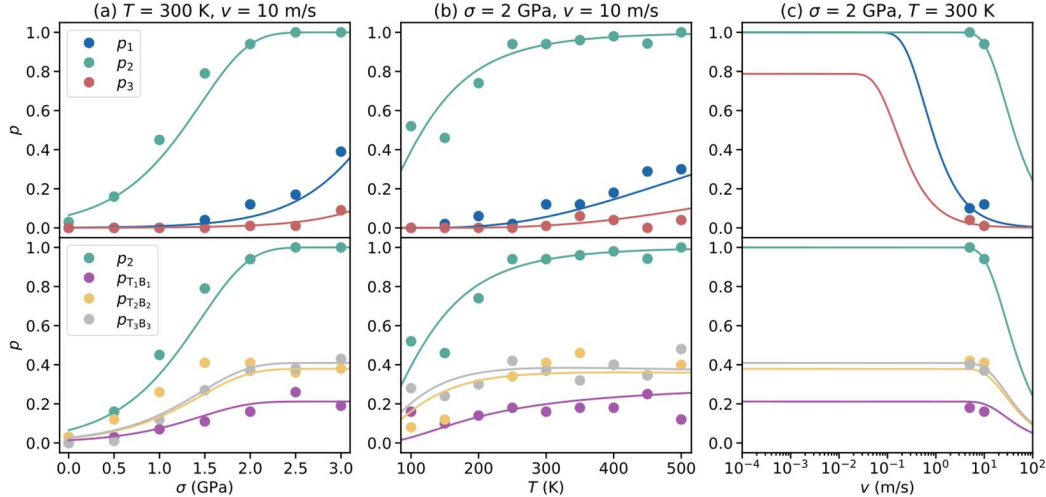

Fig. 2. Stochastic interlayer binding dynamics. Interlayer bond formation probability of various atom pairs in a 9.43° twisted $V_1V_1$ bilayer as a function of (a) external normal load, (b) temperature, and (c) sliding velocity, obtained from ensemble MD simulations (circles) and the kinetic model (Eqs. 4-7) (solid lines). In the top panels, $p_1$, $p_2$, and $p_3$ denote probabilities for the three distinct binding regions. The bottom panels present the overall probability for region 2 and the corresponding contribution of each specific binding atomic pair.

At the chosen sliding configuration, we identified five distinct interlayer bonded pairs, involving the dangling atoms of the defects. To characterize those, we mark the three dangling atoms in top layer as $T_1$, $T_2$, and $T_3$, and corresponding atoms in bottom layer as $B_1$, $B_2$, and $B_3$. Figure 1c shows a representative trajectory obtained under an external normal load of 3 GPa that incorporates three regions for consecutive bond formation and rupture events. The first event ("region 1") occurs at lateral displacement range of 2.40 nm $< x < x_1^{max} = 2.60$ nm featuring binding between the top rightmost ($T_2$) and the bottom leftmost ($B_1$) dangling atoms, which we mark as $T_2B_1$. The second event ("region 2") occurs at a lateral displacement range of 2.60 nm $< x < x_2^{max} = 2.85$ nm, featuring a $T_2B_2$ covalent bond. We note that at this eclipsed configuration, the probabilities of forming $T_1B_1$ or $T_3B_3$ bonds are comparable to that of the featured $T_2B_2$ bond, and the identity of the bond formed is determined stochastically. We further note that we did not identify any events of simultaneous multiple interlayer bond formation in this region over 100 independent trajectories. We attribute this to the fact that the simultaneous formation of two or three interlayer bonds in this region involves a significantly higher transition energy barrier (TEB) as compared to the single-bond formation (see SI section S2). Finally, the third binding event ("region 3") occurs at a lateral displacement range of 2.85 nm $< x < x_3^{max} = 3.20$ nm, demonstrating a $T_1B_2$ bonding configuration.

Bond formation in reactive sliding interfaces is a stochastic process, influenced by the external normal load, sliding velocity, and temperature, as validated through MD simulations (see Fig. 2 and SI section S5). In the top panels of Fig. 2 and SI Fig. S4, we present the total probabilities associated with the three distinct regions outlined in Fig. 1c (marked as $p_1$, $p_2$, and $p_3$), whereas in the bottom panels of the figure we show the individual bond formation probabilities of the three different pairs in region 2 (marked as $p_{T_1B_1}$, $p_{T_2B_2}$, and $p_{T_3B_3}$). All bond formation probabilities are found to grow with increasing normal load (see Fig. 2a and SI Fig. S4a), leading to the enhancement of the shear stress peak (see Fig. 1b). At a sliding velocity of $v = 10$ m/s, $p_2$ is much larger than $p_1$ and $p_3$ for all normal load considered, reaching 100% at 2.5 GPa. Specifically, in region 2, $T_2B_2$ and $T_3B_3$ pairs are more likely to bind than $T_1B_1$ with saturation probabilities of $p_{T_2B_2} \approx p_{T_3B_3} \approx 2p_{T_1B_1} \approx 40\%$. Interlayer bond formation probability increases with temperature (see Fig. 2b). In region 2, characterized by a lower TEB (see SI Fig. S2a), it saturates around room temperature, whereas in regions 1 and 3 a higher saturation temperature is observed.

The atomistic simulations presented above provide a microscopic understanding of the chemifriction mechanisms involving interlayer bond formation and rupture. Nonetheless, they are limited in their ability to reproduce experimentally relevant conditions, such as low sliding velocities (see Fig. 2c) and long sliding scenarios. To access such parameter regimes, we developed a physically motivated phenomenological kinetic model, derived based on the twisted $V_1V_1$ bilayer simulation results. The model describes the chemifriction component in terms of the rate of shear-induced interlayer bond formation and its survival probability. The former is given by the Arrhenius law (42-44):

$$r(x, \sigma, T) = f_0 e^{\left(-\frac{E_a(x) - E_m(\sigma)}{k_B T}\right)}, \qquad (1)$$

where $f_0$ is the attempt frequency, $k_B$ is the Boltzmann constant, $E_a(x)$ is the TEB in the absence of external load, and $E_m(\sigma)$ is the effective TEB reduction due to the application of normal pressure, $\sigma$. In our model, $E_a(x)$ is assumed to have a parabolic dependence on the lateral defect displacement $(x)$, obtaining a minimum $(E_{\min})$ at the eclipsed configuration (displacement of $x_e$): $E_a(x) = E_{\min} + \beta(x - x_e)^2$, where $\beta$ is a positive constant. Furthermore, the TEB reduction is assumed to be linear with normal pressure having an activation volume $\alpha$, $E_m(\sigma) = \alpha\sigma$ (45-47).

To describe the bond formation kinetics, we introduce the probability of the system to avoid interlayer bonding, $s(x, \sigma, v, T)$, up to sliding displacement, $x$, at a sliding velocity $v$, which is assumed to satisfy an irreversible first-order rate equation:

$$\frac{ds(x(t),\sigma,v,T)}{dt} = \frac{ds(x(t),\sigma,v,T)}{dx} v = -r(x, \sigma, T)\, s(x, \sigma, v, T). \tag{2}$$

Substituting Eq. 1 into Eq. 2, and solving for $s(x, \sigma, v, T)$ we obtain:

$$s(x, \sigma, v, T) = A e^{\left(-\frac{\sqrt{\pi} f_0}{2v\sqrt{\beta/(k_B T)}} e^{\left(\frac{\alpha\sigma - E_{\min}}{k_B T}\right)} \operatorname{erf}\left(\sqrt{\beta/(k_B T)}(x - x_e)\right)\right)}, \tag{3}$$

where $\operatorname{erf}(x) = \frac{2}{\sqrt{\pi}} \int_0^x e^{-t^2} dt$ is the Gaussian error function (48) and $A$ is a normalization constant, set to satisfy the initial condition $s(x = 0) = 1$, where $x = 0$ is the interlayer shift at which the lateral distance between the two defects is the largest within our laterally periodic supercell (see Fig.1a). In particular, for region 1, where only a single binding scenario is possible, Eq. 3 yields a bond formation probability of:

$$p_{T_2 B_1}(\sigma, v, T) = 1 - s_1(x_1^{max}, \sigma, v, T), \tag{4}$$

where $s_1$ is given by Eq. 3 with $E_{\min} = E_{\min}^{T_2 B_1}$ being the load-free energy barrier for $T_2 B_1$ bond formation in this region.

For region 2, where three potential binding scenarios exist, the rate term of Eq. 1 is written as a sum of the three possible bond formation rates. This, in turn, yields a total bond formation probability of (see SI section S3 for detailed derivations):

$$p_2(\sigma, v, T) = 1 - A e^{\sum_{n=1}^{3}\left(-\frac{\sqrt{\pi} f_0}{2v\sqrt{\beta/(k_B T)}} e^{\left(\frac{\alpha\sigma - E_{\min}^{T_n B_n}}{k_B T}\right)} \operatorname{erf}\left(\sqrt{\beta/(k_B T)}(x_2^{max} - x_e)\right)\right)}, \tag{5}$$

where $E_{\min}^{T_n B_n}$ is the barrier height associated with the load-free formation of a $T_n B_n$ bond, with $n = 1, 2,$ or $3$. Using this expression one can approximate the formation probability of individual interlayer pairs via:

$$p_{T_n B_n} = p_2 \frac{r_{T_n B_n}(x=x_e)}{\sum_{n=1}^{3} r_{T_n B_n}(x=x_e)}, \tag{6}$$

where the bond formation transition rate is measured at the eclipsed configuration, $x_e$, where it is maximal.

In region 3, the binding probability depends on the sliding history. If in region 2 the bond $T_1 B_1$ is formed (and eventually ruptures), healing consistently occurs by a transfer of an atom between the layers, yielding a transition from a $V_1 V_1$ to a $V_0 V_2$ configuration (pristine bottom layer and a double vacancy in the top layer). In such a case, interlayer bonding in region 3 is prevented. Hence, the probability for bond formation in this region is given by:

$$p_{T_1B_2}(\sigma, v, T) = (1 - p_{T_1B_1})(1 - s_3(x_3^{max}, \sigma, v, T)), \qquad (7)$$

where $s_3$ is given by Eq. 3 with $E_{min} = E_{min}^{T_1B_2}$ being the load-free energy barrier for $T_1B_2$ bond formation in this region.

In our calculations we chose $f_0 = 100 \text{ ns}^{-1}$(49), and set $x_e$ to be 1.52, 1.73, and 1.97 nm, corresponding to the eclipsed dangling atoms configurations in each region, respectively. The values of the rest of the kinetic model parameters are manually fit to obtain good agreement with our atomistic simulation results (see Fig. 2 and SI Fig. S4). The fitted values are $\alpha = 0.048$ eV/GPa, $\beta = 2.500$ eV/nm$^2$, and $E_{min} = 0.185, 0.125, 0.110, 0.108$, and 0.220 eV, for the $T_2B_1$, $T_1B_1$, $T_2B_2$, $T_3B_3$, and $T_1B_2$ bond pairs, respectively. These values agree well with those predicted by nudged elastic band (NEB) calculations (see SI Table S1). The chosen values for $\alpha$ and $\beta$ were further validated through zero temperature reactive sliding dynamics as detailed in SI section S4.

One can now use the developed phenomenological model to evaluate the interlayer bonding probability at experimentally relevant sliding velocities. As can be seen in Fig. 2c and SI Fig. S4c, at the low sliding velocity regime the binding probability saturates in all three regions. Specifically, under a normal load of 2 GPa, $p_1$ and $p_2$ reach the maximal value of 1 below sliding velocities of ~0.1 m/s and ~5 m/s, respectively, whereas $p_3$ saturates at 0.8 below a sliding velocity of ~0.01 m/s. We note that the latter does not reach the maximal value of 1 due to the healing process discussed above, suggesting that when a $T_1B_1$ pair forms in region 2, binding in region 3 is inhibited. As shown in SI Fig. S5, our model predicts that the saturation velocity decreases exponentially with reduction of the normal load.

Since friction in our system is dominated by interlayer chemical bonding, one can use the bonding probability functions derived above to estimate the dependence of the frictional shear stress on external parameters, such as the normal load and the sliding velocity. Marking by $f_{T_iB_j}(x)$ the shear stress trace associated with the $T_iB_j$ bonding event ($i$ and $j$ representing dangling atom indices in the top and bottom layers, respectively), the overall shear stress trace can be written as the following sum:

$$\tau(x) = p_{T_2B_1} f_{T_2B_1}(x) + \sum_{n=1}^{3} p_{T_nB_n} f_{T_nB_n}(x) + p_{T_1B_2} f_{T_1B_2}(x). \qquad (8)$$

Here, based on the atomistic simulation results, we assume that binding events in different regions are spatially separated and that in region 2 only a single binding event per trace can occur. The individual shear stress traces can be approximated using harmonic springs of effective stiffness, $k_{eff}^{T_iB_j}$, spanning from the initial bond formation position, $x_b^{T_iB_j}$, to the rupture position, $x_b^{T_iB_j} + l_r^{T_iB_j}(T, v)$, yielding:

$$f_{T_iB_j}(x) = \begin{cases} \dfrac{k_{\text{eff}}^{T_iB_j}}{A_s}\left(x - x_b^{T_iB_j}\right) & \text{for } x_b^{T_iB_j} \leq x \leq x_b^{T_iB_j} + l_r^{T_iB_j}(T,v) \\ 0 & \text{for } x < x_b^{T_iB_j} \text{ or } x > x_b^{T_iB_j} + l_r^{T_iB_j}(T,v) \end{cases}, \quad (9)$$

where $A_s$ is the in-plane surface area. Here, the bond formation position and spring stiffnesses are assumed to be constant, whereas the bond rupture lengths, $l_r(T,v)$, are assumed to depend only on the sliding velocity and the temperature (50-54). $l_r$ can be evaluated by the surface area encompassed by the bond survival probability distribution across the sliding displacement. This probability can be estimated via the same kinetic model used above (Eqs. 1-2) for bond formation, where the bond rupture energy barrier in the finite harmonic potential well decreases quadratically with lateral interlayer displacement:

$$E(x) = E_r^{max} - \frac{1}{2}k_{\text{eff}}^{T_iB_j}\left(x - x_b^{T_iB_j}\right)^2, \quad (10)$$

and $-E_r^{max}$ is the harmonic potential minimum. Substituting Eq. 10 in Eq. 1 and solving for Eq. 2, one obtains the bond survival probability (55, 56):

$$s_r(x,v,T) = e^{\dfrac{-\sqrt{\pi}f_0}{2v\sqrt{k_{\text{eff}}/(2k_BT)}} e^{\dfrac{-E_r^{max}}{k_BT}} \text{erfi}\left(\sqrt{k_{\text{eff}}/(2k_BT)}(x-x_b)\right)}, \quad (11)$$

where $\text{erfi}(x) = \frac{2}{\sqrt{\pi}}\int_0^x e^{t^2}dt$ is the imaginary error function. This allows us to evaluate the probability density function for bond rupture at a displacement interval between $x$ and $x + dx$ as $-\frac{\partial s_r(x,v,T)}{\partial x}dx$. Averaging the bond rupture length over this probability density function yields the mean bond rupture length:

$$l_r^{T_iB_j}(T,v) = \int_{x_b^{T_iB_j}}^{\infty} dx\left(x - x_b^{T_iB_j}\right)\left[-\frac{\partial s_r(x,v,T)}{\partial x}\right] = \int_{x_b^{T_iB_j}}^{\infty} dx\, s_r(x,v,T), \quad (12)$$

where we have used integration by parts and assumed that $s_r(x,v,T)$ vanishes at $x \to \infty$. This allows us to evaluate the shear stress traces, $f_{T_iB_j}(x)$, of Eq. 9 with the parameters extracted from the reactive sliding dynamics simulations (see SI Section S6 for details), which together with Eqs. 4-7 for the bond formation probabilities, $p_{T_iB_j}(\sigma,v,T)$, provide an estimation of the overall frictional shear stress traces via Eq. 8.

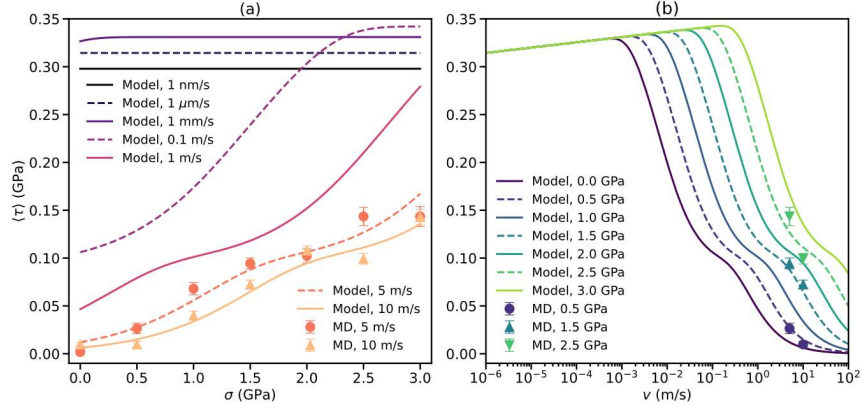

Fig. 3. Normal load (a) and velocity (b) dependence of the chemical contribution to the kinetic frictional stress of a twisted $V_1V_1$ bilayer at a temperature of 300 K. Symbols represent atomistic simulation results and lines represent model predictions (Eq. 8). Error bars for MD results are calculated from the standard error of the mean across ensembles of 100 trajectories for 10 m/s sliding velocity results and 50 trajectories for 5 m/s results. To isolate the chemifriction effect, we subtract from the atomistic simulation results the (nearly constant) frictional stress obtained for the pristine $V_0V_0$ bilayer (~0.008 GPa, see Fig. 5c of Ref. (34)), which is not considered by our phenomenological model.

The derived phenomenological model reproduces the shear stress traces obtained from the fully atomistic simulation in a broad range of normal loads with remarkable success (see Fig. 1b). Given the reliability of the model and the fact that all its assumptions and parameters should hold also outside the conditions characterizing our simulation results, we may now use it to predict chemifrictional effects in experimentally accessible sliding velocities (10, 57). Figure 3 shows the mean kinetic frictional stress dependence on the normal load (Fig. 3a) and sliding velocity (Fig. 3b) under such conditions. As expected, at the high-velocity regime there is a monotonic increase of friction with normal load, due to the corresponding increase of bond formation probability (see Fig. 2a and SI Fig. S4a). Notably, at low velocities the friction is found to be nearly independent of the applied load. This is attributed to the fact that the bond formation probability itself becomes independent of normal load at this experimentally relevant regime (see SI Fig. S6). For the velocity dependence of friction, we find two distinct regimes. At experimentally relevant velocities friction increases logarithmically. This can be rationalized by the fact that the bond rupture length increases with sliding velocity (55, 56), because thermal fluctuations do not have sufficient time to assist the bond rupture barrier crossing. A turnover is predicted to occur at higher velocities, where friction is found to drastically decrease with sliding velocity, due to reduced probability of interlayer bond formation. Notably, at the high velocity regime, both the load and the velocity dependence of the frictional stress exhibit a wavy structure. This is attributed to the difference in the TEBs associated with various interlayer bonding pairs that are activated at different velocities and normal loads.

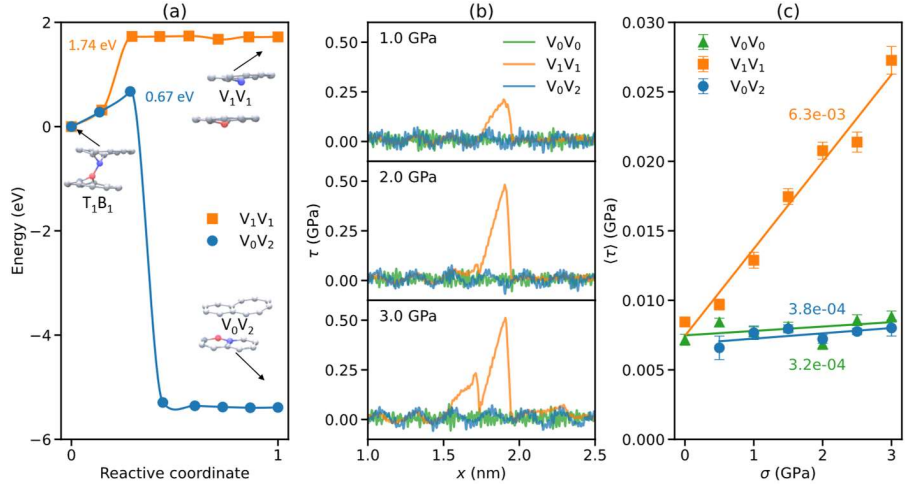

Fig. 4. Sliding induced healing. (a) NEB results for two reaction pathways of $T_1B_1$ bond rupture, one leading to the $V_1V_1$ configuration (orange) and the other to the $V_0V_2$ configuration (blue). Symbols and solid lines represent the energies of individual images and a cubic polynomial interpolation, respectively. (b) Frictional stress traces of the $V_0V_0$ (green), $V_1V_1$ (orange), and $V_0V_2$ (blue) configurations under normal loads of 1 (top sub-panel), 2 (middle sub-panel), and 3 (bottom sub-panel) GPa. (c) Average frictional stress calculated for the three configurations as a function of normal load obtained from traces such as those presented in panel (b).

As mentioned earlier, a unique sliding induced healing process occurs when the $T_1B_1$ pair bond ruptures, involving the transfer of the $T_1$ atom from the top layer to the bottom layer. This results in a $V_0V_2$ configuration, where the bottom layer is pristine, and the top layer includes a double vacancy (see the inset of Fig. 4a and SI Movie S4). NEB calculations reveal that the TEB for this healing transition (0.67 eV) is significantly lower than that for reverting back to the original $V_1V_1$ configuration (1.74 eV) upon bond rupture (see SI section S2 for computational details). The resulting $V_0V_2$ configuration presents smooth superlubric motion, similar to that of the pristine $V_0V_0$ bilayer for all normal loads considered (see Fig. 4b). At a temperature of $T = 300$ K and a sliding velocity of $v = 10$ m/s, the coefficient of friction of the $V_0V_2$ configuration is $3.2 \times 10^{-4}$, compared to $3.8 \times 10^{-4}$ for the pristine bilayer and $6.3 \times 10^{-3}$ for the $V_1V_1$ configuration, as shown in Fig. 4c. Since the healing process is initiated above a normal load of ~0.5 GPa, we expect a reduction of friction with increasing normal load around this threshold, resulting in a negative differential friction coefficient.

The revealed sliding-induced healing process can serve to constitute a run-in procedure, where pre-sliding cycles reduce friction. To demonstrate this, we conducted repeated forward and backward sliding simulations for the twisted $V_1V_1$ bilayer at $T = 300$ K and $v = 10$ m/s. To accelerate the simulations, the maximal lateral distance between two vacancies was set at 0.74 nm, corresponding to a separation of three intra-layer lattice vectors. Each trajectory consisted of 20 cycles, with a forward and backward sliding distance of 1.5 nm each. Kinetic frictional stress results were averaged over 30 independent trajectories for $\sigma = 0.5$ GPa, and 10 trajectories for $\sigma = 1.0$ and 2.0 GPa.

At a normal load of $\sigma = 1.0$ GPa and above, we find a sharp drop in the corrugation of the average friction stress traces after a few cycles ($< 5$, see Fig. 5a), corresponding to a transition from the $V_1V_1$ to the $V_0V_2$ configuration (manifested as a reduction in the survival probability of the $V_1V_1$ configuration and overall trace averaged frictional stress, see Fig. 5b). Notably, Fig. 5b also shows that the healing transition is accelerated under higher normal loads, due to the corresponding increase in the $T_1B_1$ pair interlayer bond formation probability in region 2 (see Fig. 2a). These results thus demonstrate that the suggested run-in process leads to a healing-induced transition to a superlubric state, reminiscent of that of the pristine interface, which can be further enhanced by increasing the normal load.

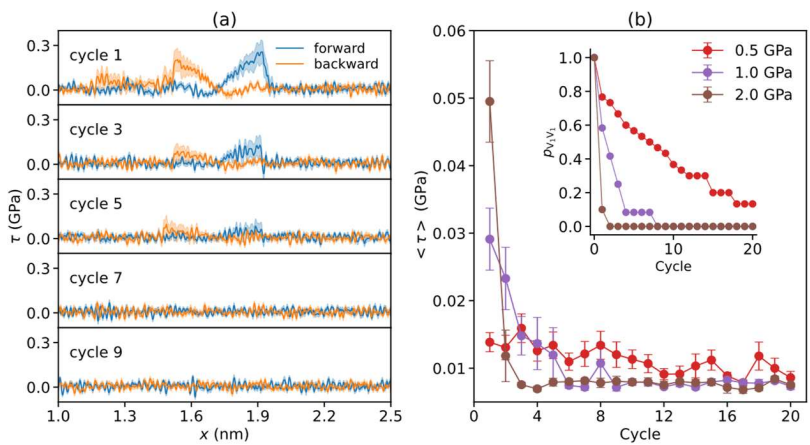

Fig. 5. Simulations of the run-in procedure. (a) Averaged friction stress traces (blue and orange lines represent the trace and retrace, respectively) obtained for nine sliding simulation cycles of a twisted bilayer starting in a $V_1V_1$ configuration under an external normal load of 1.0 GPa. Here, averaging is performed over 10 independent trajectories. The shaded regions mark the standard error of the mean. (b) The evolution of the frictional stress (main panel, averaged over the entire average trajectory) and the survival probability of the $V_1V_1$ configuration (inset, $p_{V_1V_1}$) as a function of cycle number, calculated under external normal loads of 0.5 (red), 1.0 (purple), and 2.0 (brown) GPa.

**Summary and conclusions**

In summary, using a recently developed machine learning potential trained against *ab-initio* reference data, we conducted reactive sliding dynamics simulations of a twisted incommensurate defected bilayer graphene interface. Our results reveal a unique microscopic interlayer bonding mechanism underlying *chemifriction* that may *impede* superlubric sliding. This mechanism involves consecutive stochastic bond formation and rupture events, that are spatially separated but not necessarily independent. Notably, an intricate healing process, involving interlayer atomic transfer, was identified that leads to the occurrence of negative differential friction coefficients and may be harnessed to design a friction reduction run-in procedure. Leveraging the derived atomistic insights, we developed a physically motivated phenomenological model that illustrates the interplay between bond formation kinetics, which are predominantly activated at high sliding velocities, and rupture kinetics, which become significant at lower velocities. This model allows us to predict chemifrictional effects in the experimentally accessible low sliding velocity regime, revealing a distinct transition from a logarithmic increase to a logarithmic decrease of frictional stress with velocity. While one may naively assume that chemifriction would always increase friction, our results demonstrate that under appropriate conditions, it may actually lead to the reduction of friction and *support* superlubric sliding.

## Acknowledgements

P.Y. is supported by the Israel Academy of Sciences and Humanities & Council for Higher Education Excellence Fellowship Program for International Postdoctoral Researchers. O.H. is grateful for the generous financial support of the Heineman Chair in Physical Chemistry. M.U. is grateful for generous financial support via the BSF-NSF 2023614 grant. M.U. and O.H acknowledge computational support by the Tel Aviv University Center for Nanoscience and Nanotechnology.

## References

1. K. Shinjo, M. Hirano, Dynamics of Friction - Superlubric State. *Surface Science* **283**, 473-478 (1993).
2. O. Hod, Quantifying the Stacking Registry Matching in Layered Materials. *Israel Journal of Chemistry* **50**, 506-514 (2010).
3. M. H. Müser, Structural lubricity: Role of dimension and symmetry. *Europhysics Letters (EPL)* **66**, 97-103 (2004).
4. D. Berman, A. Erdemir, A. V. Sumant, Graphene: a new emerging lubricant. *Materials Today* **17**, 31-42 (2014).
5. M. Z. Baykara, M. R. Vazirisereshk, A. Martini, Emerging superlubricity: A review of the state of the art and perspectives on future research. *Applied Physics Reviews* **5**, 041102 (2018).
6. O. Hod, E. Meyer, Q. Zheng, M. Urbakh, Structural superlubricity and ultralow friction across the length scales. *Nature* **563**, 485-492 (2018).
7. Y. Song *et al.*, Robust microscale superlubricity in graphite/hexagonal boron nitride layered heterojunctions. *Nat Mater* **17**, 894-899 (2018).
8. Y. Song, C. Qu, M. Ma, Q. Zheng, Structural Superlubricity Based on Crystalline Materials. *Small* **16**, e1903018 (2020).
9. Z. Liu *et al.*, Observation of microscale superlubricity in graphite. *Physical Review Letters* **108**, 205503 (2012).
10. J. Yang *et al.*, Observation of high-speed microscale superlubricity in graphite. *Phys Rev Lett* **110**, 255504 (2013).
11. R. Zhang *et al.*, Superlubricity in centimetres-long double-walled carbon nanotubes under ambient conditions. *Nature Nanotechnology* **8**, 912-916 (2013).
12. I. Leven, D. Krepel, O. Shemesh, O. Hod, Robust Superlubricity in Graphene/h-BN Heterojunctions. *J Phys Chem Lett* **4**, 115-120 (2013).
13. C. Androulidakis, E. N. Koukaras, G. Paterakis, G. Trakakis, C. Galiotis, Tunable macroscale structural superlubricity in two-layer graphene via strain engineering. *Nat Commun* **11**, 1595 (2020).
14. S. Zhang *et al.*, Tuning friction to a superlubric state via in-plane straining. *Proc Natl Acad Sci U S A* **116**, 24452-24456 (2019).
15. F. Banhart, J. Kotakoski, A. V. Krasheninnikov, Structural Defects in Graphene. *ACS Nano* **5**, 26-41 (2011).
16. P. Gajurel *et al.*, Vacancy‐Controlled Contact Friction in Graphene. *Advanced Functional Materials* **27**, 1702832 (2017).
17. A. Zambudio *et al.*, Fine defect engineering of graphene friction. *Carbon* **182**, 735-741 (2021).

18. Z. Chen, A. Khajeh, A. Martini, S. H. Kim, Chemical and physical origins of friction on surfaces with atomic steps. *Sci Adv* **5**, eaaw0513 (2019).
19. K. Wang, C. Qu, J. Wang, B. Quan, Q. Zheng, Characterization of a Microscale Superlubric Graphite Interface. *Phys Rev Lett* **125**, 026101 (2020).
20. M. M. van Wijk, M. Dienwiebel, J. W. M. Frenken, A. Fasolino, Superlubric to stick-slip sliding of incommensurate graphene flakes on graphite. *Physical Review B* **88** (2013).
21. M. Liao *et al.*, Ultra-low friction and edge-pinning effect in large-lattice-mismatch van der Waals heterostructures. *Nature Materials* **21**, 47-53 (2022).
22. Y. Tison *et al.*, Grain boundaries in graphene on SiC(000$\bar{1}$) substrate. *Nano Lett* **14**, 6382-6386 (2014).
23. J. H. Jeong, Y. Jung, J. U. Park, G. H. Lee, Gate-Tunable Electrostatic Friction of Grain Boundary in Chemical-Vapor-Deposited MoS(2). *Nano Lett* **23**, 3085-3089 (2023).
24. A. Vanossi, N. Manini, M. Urbakh, S. Zapperi, E. Tosatti, Colloquium: Modeling friction: From nanoscale to mesoscale. *Reviews of Modern Physics* **85**, 529-552 (2013).
25. S. Li *et al.*, The evolving quality of frictional contact with graphene. *Nature* **539**, 541-545 (2016).
26. B. Shen *et al.*, Double-Vacancy Controlled Friction on Graphene: The Enhancement of Atomic Pinning. *Langmuir* **35**, 12898-12907 (2019).
27. X. Gao, W. Ouyang, M. Urbakh, O. Hod, Superlubric polycrystalline graphene interfaces. *Nature Communications* **12**, 5694 (2021).
28. A. S. Minkin, I. V. Lebedeva, A. M. Popov, A. A. Knizhnik, Atomic-scale defects restricting structural superlubricity: Ab initio study on the example of the twisted graphene bilayer. *Physical Review B* **104** (2021).
29. X. Gao, M. Urbakh, O. Hod, Stick-Slip Dynamics of Moire Superstructures in Polycrystalline 2D Material Interfaces. *Physical Review Letters* **129**, 276101 (2022).
30. Y. Song *et al.*, Velocity Dependence of Moire Friction. *Nano Lett* **22**, 9529-9536 (2022).
31. L. Zhang *et al.*, Nonmonotonic Effects of Atomic Vacancy Defects on Friction. *ACS Applied Materials & Interfaces* **15**, 45455-45464 (2023).
32. J. Wang, A. Khosravi, A. Vanossi, E. Tosatti, Colloquium: Sliding and pinning in structurally lubric 2D material interfaces. *Reviews of Modern Physics* **96** (2024).
33. W. Yan *et al.*, Shape-dependent friction scaling laws in twisted layered material interfaces. *Journal of the Mechanics and Physics of Solids* **185** (2024).
34. P. Ying, A. Natan, O. Hod, M. Urbakh, Effect of Interlayer Bonding on Superlubric Sliding of Graphene Contacts: A Machine-Learning Potential Study. *ACS Nano* **18**, 10133-10141 (2024).
35. M. H. Gass *et al.*, Free-standing graphene at atomic resolution. *Nat Nanotechnol* **3**, 676-681 (2008).
36. J. C. Meyer *et al.*, Direct Imaging of Lattice Atoms and Topological Defects in Graphene Membranes. *Nano Letters* **8**, 3582-3586 (2008).
37. M. M. Ugeda, I. Brihuega, F. Guinea, J. M. Gomez-Rodriguez, Missing atom as a source of carbon magnetism. *Phys Rev Lett* **104**, 096804 (2010).
38. Y. Zhang *et al.*, Scanning Tunneling Microscopy of the π Magnetism of a Single Carbon Vacancy in Graphene. *Physical Review Letters* **117**, 166801 (2016).
39. A. P. Thompson *et al.*, LAMMPS-a flexible simulation tool for particle-based materials modeling at the atomic, meso, and continuum scales. *Computer Physics Communications* **271**, 108171 (2022).
40. R. A. Bernal *et al.*, Influence of chemical bonding on the variability of diamond-like carbon nanoscale adhesion. *Carbon* **128**, 267-276 (2018).
41. A. Stukowski, Visualization and analysis of atomistic simulation data with OVITO–the Open Visualization Tool. *Modelling and Simulation in Materials Science and Engineering* **18**, 015012 (2010).
42. A. Martini, S. H. Kim, Activation Volume in Shear-Driven Chemical Reactions. *Tribology Letters* **69**, 150 (2021).
43. H. Spikes, Stress-augmented thermal activation: Tribology feels the force. *Friction* **6**, 1-31 (2018).
44. T. D. B. Jacobs, B. Gotsmann, M. A. Lantz, R. W. Carpick, On the Application of Transition State Theory to Atomic-Scale Wear. *Tribology Letters* **39**, 257-271 (2010).
45. Y. S. Zholdassov *et al.*, Acceleration of Diels-Alder reactions by mechanical distortion. *Science* **380**, 1053-1058 (2023).
46. F. H. Bhuiyan, Y. S. Li, S. H. Kim, A. Martini, Shear-activation of mechanochemical reactions through molecular deformation. *Scientific Reports* **14**, 2992 (2024).
47. C. Tang *et al.*, Graphene Failure under MPa: Nanowear of Step Edges Initiated by Interfacial Mechanochemical Reactions. *Nano Lett* **24**, 3866-3873 (2024).
48. L. C. Andrews, *Special functions of mathematics for engineers* (Spie Press, 1998), vol. 49.
49. B. N. Persson, *Sliding friction: physical principles and applications* (Springer Science & Business Media, 2013).
50. A. E. Filippov, J. Klafter, M. Urbakh, Friction through dynamical formation and rupture of molecular bonds. *Physical Review Letters* **92**, 135503 (2004).
51. H. Y. Chen, Y. P. Chu, Theoretical determination of the strength of soft noncovalent molecular bonds. *Physical Review E* **71**, 010901 (2005).

52. I. Barel, M. Urbakh, L. Jansen, A. Schirmeisen, Multibond Dynamics of Nanoscale Friction: The Role of Temperature. *Physical Review Letters* **104** (2010).
53. R. W. Friddle, A. Noy, J. J. De Yoreo, Interpreting the widespread nonlinear force spectra of intermolecular bonds. *Proceedings of the National Academy of Sciences* **109**, 13573-13578 (2012).
54. D. Li, B. Ji, Predicted rupture force of a single molecular bond becomes rate independent at ultralow loading rates. *Physical Review Letters* **112**, 078302 (2014).
55. G. Hummer, A. Szabo, Kinetics from Nonequilibrium Single-Molecule Pulling Experiments. *Biophysical Journal* **85**, 5-15 (2003).
56. O. K. Dudko, A. E. Filippov, J. Klafter, M. Urbakh, Beyond the conventional description of dynamic force spectroscopy of adhesion bonds. *Proceedings of the National Academy of Sciences* **100**, 11378-11381 (2003).
57. M. Dienwiebel *et al.*, Superlubricity of graphite. *Phys Rev Lett* **92**, 126101 (2004).

# Supplementary Information

# Chemifriction and Superlubricity: Friends or Foes?

Penghua Ying[1], Xiang Gao[1], Amir Natan,[2] Michael Urbakh[1], Oded Hod[1]

[1] *Department of Physical Chemistry, School of Chemistry, The Raymond and Beverly Sackler Faculty of Exact Sciences and The Sackler Center for Computational Molecular and Materials Science, Tel Aviv University, Tel Aviv 6997801, Israel*

[2] *Department of Physical Electronics, Tel-Aviv University, Tel-Aviv, 6997801, Israel*

This supporting information document includes the following sections:

1. Reactive sliding dynamics simulation setup
2. Nudged elastic band calculations
3. Bond formation probability in region 2
4. Reactive sliding dynamics at zero temperature
5. Additional reactive sliding dynamics simulations at finite temperature
6. Parameters for estimating bond rupture length

# 1. Reactive sliding dynamics simulation setup

The twisted $V_1V_1$ bilayer model system was constructed following the method described in Ref. (1) (see Fig. 1 of the main text). Periodic boundary conditions were employed in the in-plane directions with lateral dimensions of $5.18 \times 1.50$ nm$^2$, and a vacuum layer of $\sim 15$ Å was used in the out-of-plane direction. These supercell dimensions were found to be sufficiently large to avoid interactions between adjacent defect images.

During the reactive sliding simulations the top layer was laterally driven along the zigzag ($x$-axis) graphene lattice direction by a rigid slider (modeled by a rigid pristine graphene layer) with a spring of stiffness $K_1 = 50$ N/m, and the bottom layer was anchored with harmonic springs of the same stiffness, $K_2 = 50$ N/m, to their original positions, effectively simulating a double-layer substrate (see Fig. S1 and SI Section 4 of Ref. (2)). In each trajectory, unless otherwise specified, a total relative sliding displacement of 5.18 nm was implemented. The instantaneous frictional stress trace was calculated as $\tau = K_1(x - X_{com})/A_s$, where $X_{com}$ is the center of mass $x$ coordinate of the top layer, $x = vt$ is the displacement of the rigid layer, and $A_s$ is the contact area, defined as the interface area of the $V_1V_1$ bilayer supercell. For each trajectory, the kinetic friction was then determined as the mean shear stress $\langle \tau \rangle$ over the entire trace. To evaluate the bond formation and rupture probabilities we captured 518 atomic snapshots, corresponding to displacement intervals of 0.1 Å. For each snapshot, the binding state of each dangling pair was determined as bonded or non-bonded based on a cutoff distance of 1.8 Å.

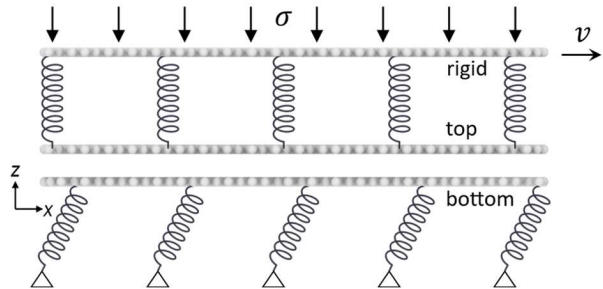

Fig. S1: A schematic representation of the simulation setup.

We consider the influences of sliding velocity, $v$, external normal load, $\sigma$, and temperature, $T$, on the sliding friction of the reactive sliding interfaces. The sliding velocity was imposed by moving the rigid slider at a target constant value; the external normal pressures was applied by exerting a vertical force to each atom of the rigid slider (see Fig. S1); and temperature was set by using a Langevin thermostat (3) applied to all sliding interface flexible atoms in all three direction with a damping parameter of $1$ ps. Given the stochastic nature of interlayer bonding kinetics, we considered 100 independent sliding trajectories (each with a different random seed of the initial atomic velocity distribution) to calculate the average friction trace and corresponding kinetic friction for the $V_1V_1$ bilayer at $T = 300$ K and $v = 10$ m/s; and 50 trajectories in all other cases investigated. The kinetic friction stress was evaluated from the average of these independent trajectories, and the standard error was calculated to obtain error bars.

## 2. Nudged elastic band calculations

To evaluate the transition energy barrier (TEB) for interlayer bond formation of different atomic pairs, we used the nudged elastic band (NEB) method (5), as implemented in the Atomic Simulation Environment (ASE) package (6). Atomic forces were calculated using ASE, interfaced with the developed NequIP model (7, 8). Atomic position optimizations for the initial (reactants) and final (products) structures were performed using the FIRE algorithm (9) with a fixed supercell box and an atomic force convergence criterion of $0.01$ eV/Å. Eighteen interior images were interpolated between the initial and final optimized structures to establish a preliminary reaction path. All replicas were connected by springs of stiffness of $0.1$ eV/Å$^2$ and the energy of the entire band was minimized using the FIRE algorithm with a force tolerance of $0.02$ eV/Å to determine the reaction path and the TEB values (see Fig. S2).

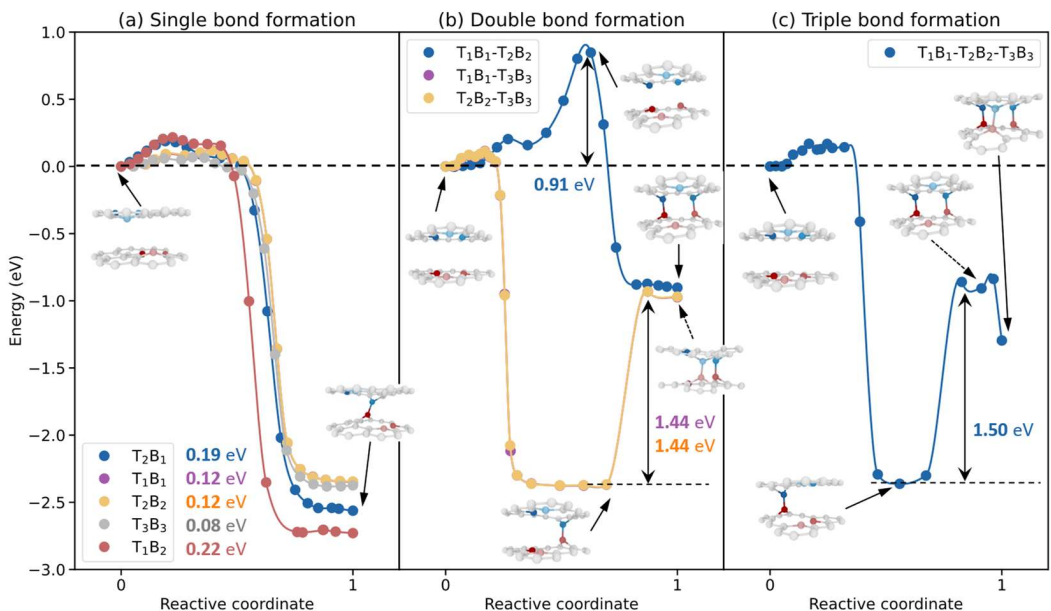

Fig. S2. NEB predictions of the reaction paths and TEBs for different interlayer bond formation scenarios: (a) single bond formation of five different atomic pairs across the three displacement regions; (b) simultaneous double bond formation in region 2 for three different atomic pairs; and (c) simultaneous triple bond formation in region 2. The insets in (a-c) are representative snapshots along the reaction path.

Figure S2a displays the reaction paths for single interlayer bond formation of five different atomic pairs (see Fig. 1c of the main text). The $T_2B_1$ (region 1) and $T_1B_2$ (region 3) bond formation energy barriers (estimated as the energy difference between the saddle-point and the unbound state) of 0.19 and 0.22 eV, respectively, are found to be considerably larger than the corresponding barriers (~0.1 eV) for the other three pairs, i.e., $T_1B_1$, $T_2B_2$, and $T_3B_3$ in region 2. As can be seen in Table S1, the interlayer bond formation TEB values obtained using the atomistic simulations are well reproduced by the kinetic model. The bond rupture TEBs (estimated as the energy difference between the saddle-point and the bonded state) are significantly larger, lying in the range of $2.5 - 3.0$ eV, which is on the order of the values extracted from MD simulations (see Table S2).

As mentioned in the main text, in region 2 we observed that only a single covalent interlayer bond forms throughout all MD trajectories. This phenomenon is attributed to the significantly higher TEBs associated with the simultaneous formation of double or triple bonds, as demonstrated in Figs. S2b and S2c. For instance, the consecutive formation of $T_1B_1$ and $T_2B_2$ interlayer bonds involves a TEB of 0.91 eV, whereas for the sequential formations of the $T_1B_1$ and $T_3B_3$ bond pair and of the $T_2B_2$ and $T_3B_3$ bond pair, the second bond formation is associated with a TEB of 1.44 eV (see Fig. S2b). Similarly, the formation of three bonds involves a TEB of 1.50 eV for the second binding event. Notably, even in this case, forming the third bond requires overcoming a much lower TEB, ~0.1 eV (see Fig. S2c), exemplifying that the formation of the second bond poses the primary challenge in the eventual formation of two or three interlayer bonds.

Table S1. TEBs for interlayer single bond formation involving five different atomic pairs, as predicted by NEB calculations and by the kinetic model.

| Method | TEBs (eV) | | | | |
| --- | --- | --- | --- | --- | --- |
| | $T_2B_1$ | $T_1B_1$ | $T_2B_2$ | $T_3B_3$ | $T_1B_2$ |
| NEB | 0.190 | 0.121 | 0.118 | 0.073 | 0.218 |
| kinetic model | 0.185 | 0.125 | 0.110 | 0.108 | 0.220 |

## 3. Bond formation probability in region 2

As noted in the main text, interlayer bond formation kinetics in region 2 involves three different atomic pairs: $T_1B_1$, $T_2B_2$, and $T_3B_3$. Assuming that the three binding events are independent, the kinetic equation for the probability to be in the unbound state (Eq. 2 of the main text) is generalized as follows:

$$\frac{ds_2}{dt} = -s_2 \sum_{n=1}^{3} r_{T_nB_n}, \quad (S1)$$

where $r_{T_nB_n}$ is the formation rate of atomic pair $T_nB_n$ ($n = 1, 2, 3$). This additive form is valid as the probability of forming second and third bonds (simultaneously or in parallel) with the first one is extremely low according to our simulations (see also SI section 2 above).

Incorporating the expression for $r_{T_nB_n}$ from Eq. 1 in the main text into Eq. S1 yields:

$$\frac{ds_2}{dx} = \frac{-f_0 s_2}{v} \sum_{n=1}^{3} \exp\left(\frac{\alpha\sigma - \Delta E_{min}^{T_nB_n} - \beta(x-x_e)^2}{k_B T}\right), \quad (S2)$$

where $\Delta E_{min}^{T_nB_n}$ denotes the corresponding load-free bond formation energy barrier. Solving this equation with the initial condition $s_2(x = 0) = 1$ (where the origin is assumed to be sufficiently far from the eclipsed configuration) yields:

$$s_2(x, \sigma, v, T) = A \cdot \text{Exp}\left\{\sum_{n=1}^{3}\left[-\frac{\sqrt{\pi}f_0}{2v\sqrt{\beta/(k_B T)}} e^{\left(\frac{\alpha\sigma - E_{min}^{T_nB_n}}{k_B T}\right)} \text{erf}\left(\sqrt{\beta/(k_B T)}(x - x_e)\right)\right]\right\}$$

(S3)

Following Eq. 4 of the main text, the bond formation probability in region 2, $p_2$, can be obtained from $p_2(\sigma, v, T) = 1 - s_2(x_{max}, \sigma, v, T)$, yielding Eq. 5 of the main text.

## 4. Reactive sliding dynamics at zero temperature

Inspired by the two-state model introduced in Ref. (10), the zero temperature friction stress trace can be estimated as follows:

$$\tau = \frac{k_{\text{eff}}}{A_s}(x - x_b)\big[1 - H\big(E_b(\sigma, x)\big)\big]H\big(E_r(x)\big), \tag{S4}$$

where the bond formation TEB is given by $E_b(\sigma, x) = E_{min} + \beta(x - x_e)^2 - \alpha\sigma$, the bond rupture TEB reads as $E_r(x) = E_r^{max} - \frac{1}{2}k_{\text{eff}}(x - x_b)^2$, and $A_s$ is the contact area. Here, the Heaviside step function, $H$, defines the threshold conditions for interlayer bond formation (when the corresponding barrier nullifies $E_b(\sigma, x) = 0$) and rupture (when $E_r(x) = 0$).

To validate Eq. S4, we conducted reactive sliding dynamics simulations for the twisted $V_1V_1$ bilayer at zero temperature ($T = 0$ K) and a sliding velocity of 10 m/s under different external normal pressures. Using the parameters appearing in Tables S1 and S2, Eq. S7 accurately reproduces the atomistic simulation results (see Fig. S3). Below 2.5 GPa, bond formation does not occur, leading to ultralow friction throughout the trajectory (see top sub-panels of Figs. S3a, b). Conversely, at or above 2.5 GPa, bond formation of the $T_3B_3$ pair is observed in region 2, generating a distinct peak around an interlayer displacement of 2.6 nm (see the lower subpanels of Figs. S3a, b), which is manifested as an abrupt increase of the kinetic friction (see Fig. S3c).

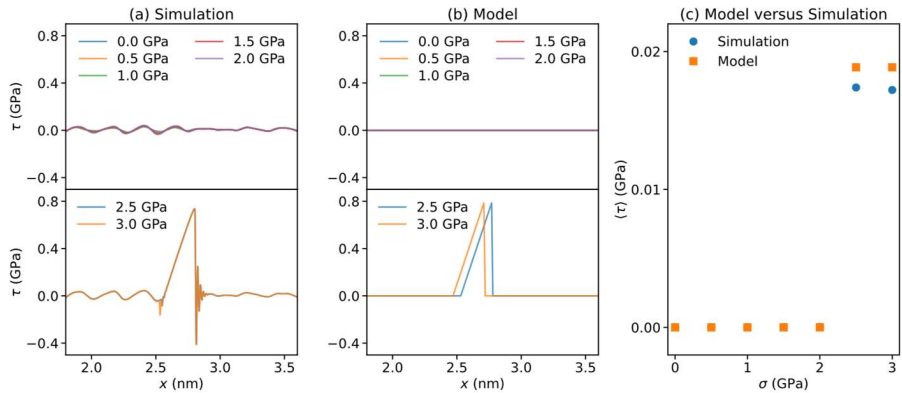

Fig. S3. Reactive sliding dynamics at zero temperature ($T = 0$ K). The friction trace of a twisted $V_1V_1$ bilayer under different external normal loads obtained from (a) atomistic MD simulation and (b) the model (Eq. S4).

Panel (c) presents the kinetic friction (evaluated as the force average over the entire trace) as a function of normal load, comparing results from MD simulation (blue circles) with those derived from Eq. S4 (orange squares).

## 5. Additional reactive sliding dynamics simulations at finite temperature

Following the simulation results presented in Fig. 2 of main text, we conducted additional reactive sliding dynamics simulations at finite temperature to further validate our stochastic model for bond formation probability (Eqs. 4-7 of the main text). To that end, we performed simulations under varying external normal loads (see Fig. S4a), temperatures (see Fig. S4b), and sliding velocities (see Fig. S4c). Overall, the simulation results present the same qualitative picture as that seen in the main text and the kinetic model captures them well. Some quantitative variations are observed due to the different sliding conditions. For example, the transition pressure appearing in panel (a) downshifts with respect to that appearing in the main text, since the lower sliding velocity allows for more time for bond formation to occur.

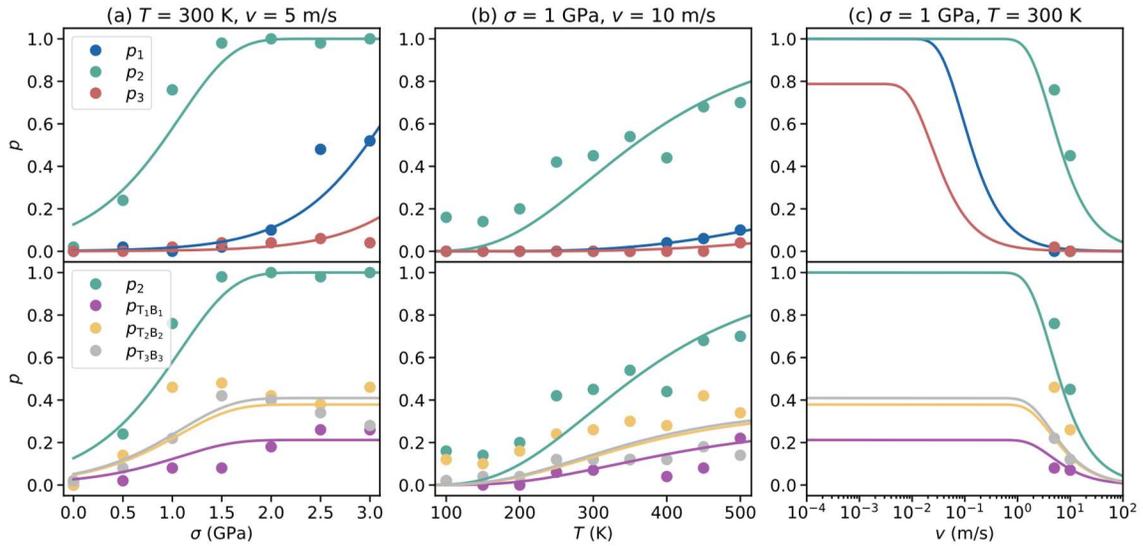

Fig. S4. Additional stochastic interlayer binding dynamics results. Interlayer bond formation probability of various atom pairs in a 9.43° twisted $V_1V_1$ bilayer as a function of (a) external normal load, (b) temperature, and (c) sliding velocity, obtained from ensemble MD simulations (circles) and the kinetic model (solid lines). The top panels show the probabilities for the three distinct binding regions, and the bottom panels present the overall probability for region 2 and the corresponding contribution of each specific binding atomic pair.

Similarly, the kinetic model predicts that the velocity, under which the bond formation probabilities saturate, downshifts with a reduction of the normal load (see Fig. S5a) and that the dependence is exponential (see Fig. S5b).

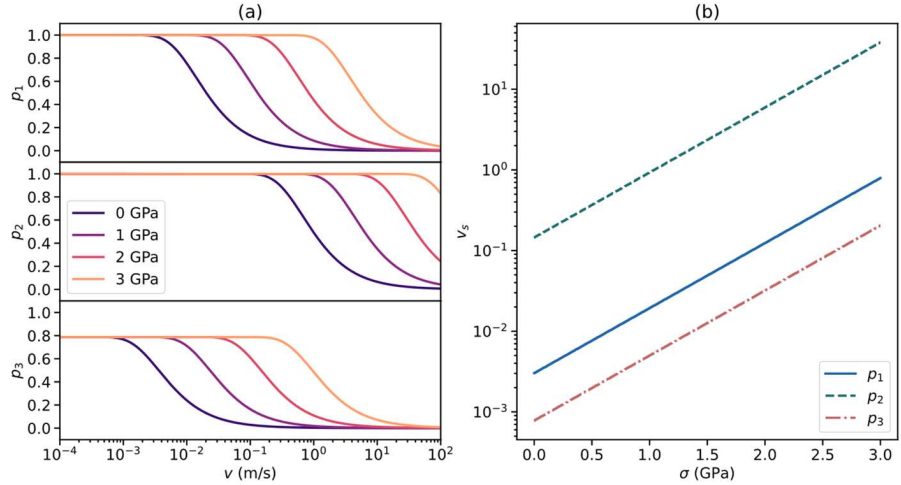

Fig. S5. (a) Interlayer bond formation probabilities as a function of sliding velocity, under various external normal loads, predicted by the kinetic model for regions 1 ($p_1$, top subpanel), 2 ($p_2$, middle subpanel), and 3 ($p_3$, lower subpanel). (b) Load dependence of the velocity under which the bond formation probability saturates, $v_s$, for the three sliding regions.

Finally, to explain the load-independent frictional stress observed in Fig. 3a of the main text, in Fig. S6 we use the kinetic model to plot the interlayer bond formation probabilities for the three regions as a function of normal load. Indeed, at experimentally relevant sliding velocities of $\lesssim 1\,\mathrm{mm/s}$, all interlayer bond formation probabilities, which dominate friction, are found to be independent of the normal load.

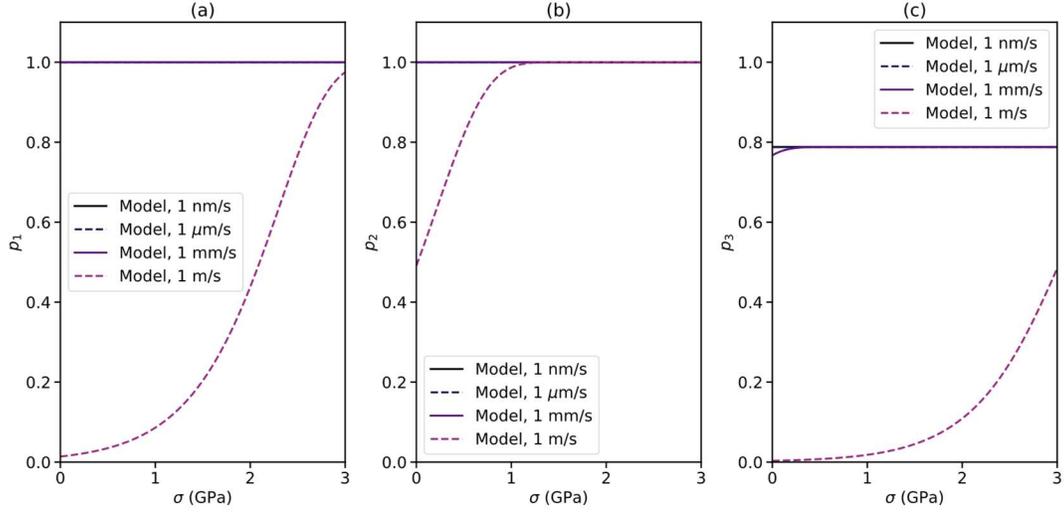

Fig. S6. Interlayer bond formation probability in (a) region 1, (b) region 2, and (c) region 3 for a 9.43° twisted $V_1V_1$ bilayer as a function of external normal load, as predicted by the kinetic model. For all three regions, four representative sliding velocities, i.e., 1 nm/s, 1 µm/s, 1 mm/s, and 1 m/s are considered.

## 6. Parameters for estimating bond rupture length

In this section, we discuss the choice of parameters used for calculating the bond rupture length via Eq. 12 of the main text. In Fig. S7 we show the frictional stress traces for each bonded atomic pair obtained from reactive sliding dynamics simulations of the twisted $V_1V_1$ bilayer at a temperature of $T = 300$ K and a sliding velocity of $v = 10$ m/s, under three normal loads. For each atomic pair, we considered the friction trace section spanning the region between bond formation and rupture positions, as determined by observing snapshots along the trace. Based on these simulation traces, Fig. S8 shows the average bond formation position, $x_b$, bond rupture length, $l_r$, and effective bond stiffness, $k_{\text{eff}}$ (obtained via linear fits of the traces appearing in Fig. S7), as functions of normal load. All three parameters for all pairs appear to be unaffected by the external normal load, in the considered range. Therefore, in Eq. 9 of the main text, $x_b$, $l_r$, and $k_{\text{eff}}$ are taken to be load independent. The bond rupture TEB can then be estimated using the spring energy equation, $E_r^{\max} = \frac{1}{2} k_{\text{eff}} l_r^2$. We note that the obtained values of $E_r^{\max}$ are significantly higher than the thermal energy (see Table S2), such that extracting the value of $l_r$ from simulations performed at room temperature is a well justified approximation. All values for the abovementioned kinetic model bond rupture parameters appear in Table S2.

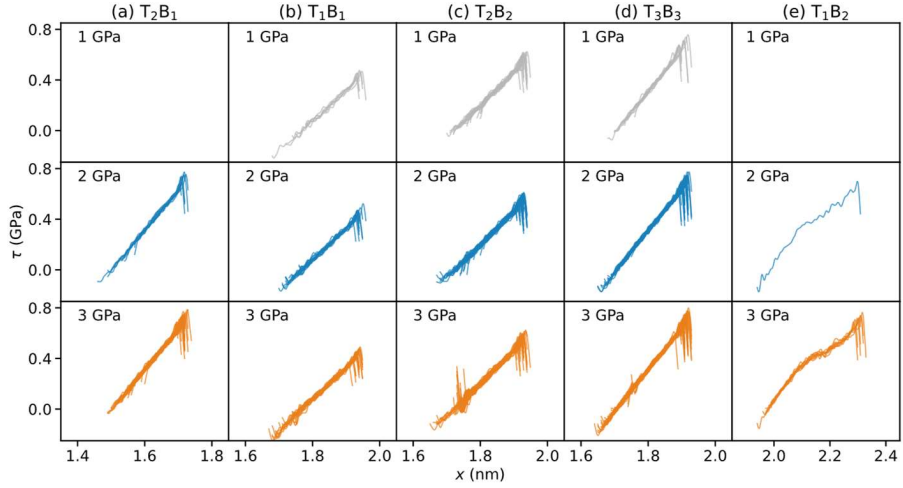

Fig. S7. Frictional shear stress traces for five bonded pairs: (a) $T_2B_1$; (b) $T_1B_1$; (c) $T_2B_2$; (d) $T_3B_3$; and (e) $T_1B_2$; obtained from reactive sliding dynamics simulation at a temperature of $T = 300$ K, a sliding velocity of $v = 10$ m/s, and under normal loads of $\sigma = 1$ GPa (top panels), 2 GPa (middle panels), and 3 GPa (bottom panels). The $T_2B_1$ and $T_1B_2$ traces under 1 GPa are not depicted as no interlayer bond formation was found to occur for these atomic pairs under the considered conditions.

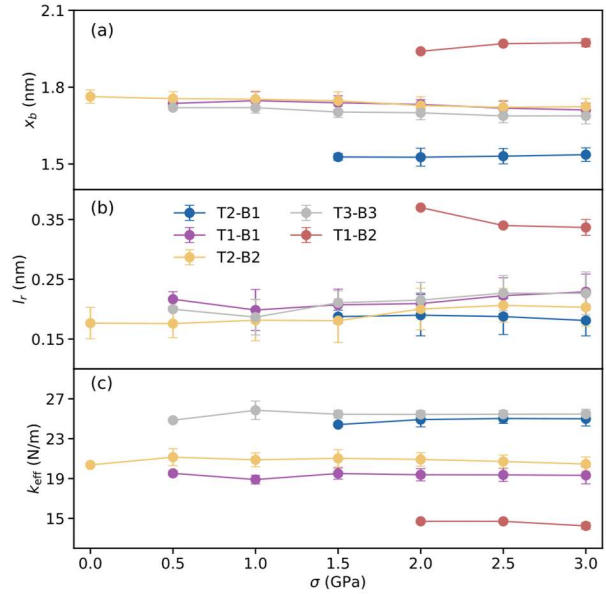

Fig. S8. Demonstration of the independence of (a) bond formation position ($x_b$), (b) rupture length ($l_r$), and (c) effective stiffness ($k_{\text{eff}}$) on the normal load ($\sigma$), as obtained from reactive sliding dynamics simulations at a temperature of $T = 300$ K and a sliding velocity of $v = 10$ m/s.

Table S2. Kinetic model interlayer bond rupture parameters (bond formation position ($x_b$), bond rupture length ($l_r$), load-free TEB ($E_r^{\max}$), and effective bond stiffness ($k_{\text{eff}}$)) for five bonded pairs.

| Bonded pairs | $T_2B_1$ | $T_1B_1$ | $T_2B_2$ | $T_3B_3$ | $T_1B_2$ |
| --- | --- | --- | --- | --- | --- |
| $x_b$ (nm) | 1.53 | 1.73 | 1.74 | 1.70 | 1.96 |
| $l_r$ (Å) | 1.9 | 2.1 | 1.9 | 2.1 | 3.5 |
| $E_r^{\max}$ (eV) | 2.70 | 2.76 | 2.32 | 3.53 | 5.53 |
| $k_{\text{eff}}$ (N/m) | 24.8 | 19.3 | 20.8 | 25.4 | 14.6 |

# References

1. A. R. Muniz, D. Maroudas, Opening and tuning of band gap by the formation of diamond superlattices in twisted bilayer graphene. *Physical Review B* **86**, 075404 (2012).

2. P. Ying, A. Natan, O. Hod, M. Urbakh, Effect of Interlayer Bonding on Superlubric Sliding of Graphene Contacts: A Machine-Learning Potential Study. *ACS Nano* **18**, 10133-10141 (2024).

3. T. Schneider, E. Stoll, Molecular-dynamics study of a three-dimensional one-component model for distortive phase transitions. *Physical Review B* **17**, 1302 (1978).

4. A. Stukowski, Visualization and analysis of atomistic simulation data with OVITO–the Open Visualization Tool. *Modelling and Simulation in Materials Science and Engineering* **18**, 015012 (2010).

5. H. Jónsson, G. Mills, K. W. Jacobsen, "Nudged elastic band method for finding minimum energy paths of transitions" in Classical and quantum dynamics in condensed phase simulations. (World Scientific, 1998), pp. 385-404.

6. A. H. Larsen *et al.*, The atomic simulation environment—a Python library for working with atoms. *Journal of Physics: Condensed Matter* **29**, 273002 (2017).

7. https://github.com/mir-group/nequip.

8. http://doi.org/10.5281/zenodo.10374206.

9. E. Bitzek, P. Koskinen, F. Gähler, M. Moseler, P. Gumbsch, Structural relaxation made simple. *Physical Review Letters* **97**, 170201 (2006).

10. X. Gao, W. Ouyang, M. Urbakh, O. Hod, Superlubric polycrystalline graphene interfaces. *Nature Communications* **12**, 5694 (2021).